\documentclass[11pt,aps,showpacs,preprintnumbers,amsmath,amssymb,nofootinbib,preprint]{article}

\usepackage{epsfig}
\usepackage{dcolumn}
\usepackage{amsmath}
\usepackage{color}
\usepackage{graphicx}

\newcommand{\ba} {\begin{eqnarray}}
\newcommand{\ea} {\end{eqnarray}}
\def \be  {\begin{equation}}
\def \ee  {\end{equation}}
\def \ee  {\end{equation}}
\def \bea {\begin{eqnarray}}
\def \eea {\end{eqnarray}}

\newcommand{\nn}{\nonumber}

\usepackage{graphicx}
\usepackage{dcolumn} 
\usepackage{bm}      

\begin{document}

\title{\vspace*{-1cm}\hfill {\small\bf ECTP-2016-03} \\ \hfill {\small\bf WLCAPP-2016-03}\\
\vspace*{2cm}
Perturbative instability of inflationary cosmology from quantum potentials}

\author{A~TAWFIK\footnote{Corresponding author: a.tawfik@eng.mti.edu.eg} 
~and~A~DIAB  \\
{\it Egyptian Center for Theoretical Physics (ECTP),}\\
{\it Modern University for Technology and Information (MTI), 11571 Cairo, Egypt} and\\
{\it World Laboratory for Cosmology And Particle Physics (WLCAPP), Cairo, Egypt. }  
\\ \\
E Abou El Dahab\\
{\it Faculty of Computer Science, Modern University for Technology and}\\
{\it   Information (MTI), 11571 Cairo, Egypt}\\
{\it World Laboratory for Cosmology And Particle Physics (WLCAPP), Cairo, Egypt}
}
\date{\today}
\maketitle

\begin{abstract}

It was argued that the Raychaudhuri equation with a quantum correction term seems to avoid the Big Bang singularity and to characterize an everlasting Universe [PLB741,276(2015)]. Critical comments on both conclusions and on the correctness of the key expressions of this work were discussed in literature [MPLA31(2016)1650044]. In the present work, we have analyzed the perturbative (in)stability conditions in the inflationary era of the early Universe. We conclude that both unstable and stable modes are incompatible with the corresponding ones obtained in the standard FLRW Universe. We have shown that unstable modes do exist at small (an)isotropic perturbation and for different equations of state. Inequalities for both unstable and stable solutions with the standard FLRW space were derived. They reveal that in the FLRW flat Universe both perturbative instability and stability are likely. While negative stability modes have been obtained for radiation- and matter-dominated eras, merely, instability modes exist in case of a finite cosmological constant and also if the vacuum energy dominates the cosmic background geometry.  

\end{abstract}

\\ 
{\bf \hspace*{0.45cm} PACS Nos:}~98.80.Cq, 04.20.-q, 04.20.Cv, 98.80.Cq


\section{Introduction \label{sec:intr}}


Well approved cosmological models assume that our physical Universe is homogeneous and isotropic \cite{Fixsen:1996,Gulkis:1990s}; widely known as Friedmann-Lem$\hat{\mbox{a}}$itre-Robertson-Walker (FLRW) space \cite{Weinberg1972,Misner:1973,Misner:1984,Kolb:1990a}. Various cosmological problems arising with this standard model for cosmology, such as flatness, monopole and event horizon, are conjectured to be solved when imposing an inflation era to the early stages of the Universe \cite{Guth:1981a,Kolb1983,Accetta:1898,Kao:2007t}. Different inflationary potentials endorse successful ideas enabling us to explain recent astrophysical observations, see for instance \cite{Starobinsky:1979et, Guth:1981a, Sato:1981se,Starobinsky:1982qw,Hawking:1982sw,Guth:1982uy, Linde:1982sq,Albrecht:1982,Linde:2002q,Liddle:2003w}. 

A careful investigation of the FLRW Universe (in)stability against spatially homogeneous and isotropic perturbations is very crucial in order to legitimate the necessity, the relevance and the correctness of the proposed cosmological models. In a recent example \cite{Ali:2015}, quantum potential corrections to the FLRW Universe have been imposed. It was concluded that these {\it ad hoc} corrections avoid the Big Bang singularity and propose that the age of the resulting Universe becomes infinite. Severe criticisms on the correctness of this approach have been discussed in Ref. \cite{Lashin:2015,Lashin:2016}. It was argued that both conclusions (absence of the Big Bang singularity and the infinite age of the Universe) are badly wrong. Considerable improvements to the correction term, itself, which as the name says includes the entire quantum corrections, have been introduced \cite{Lashin:2015}. Based on our systematic analysis of the perturbative stability of both versions (\cite{Ali:2015,Lashin:2015}), we have concluded that the proposed quantum potentials apparently worsen the stability of the physical Universe \cite{Tawfik:2016biw}.

It obvious that any stable mode represents an isotropically expanding solution. This solution should be stable when an anisotropic perturbation is added to \cite{Dobado:1993a,Dobado:1995,Kao:1991,Kao:2001a,Kao:2001b,Kao:2010q}. In fact, the physical Universe could be anisotropic in the very early stages of its evolutionary eras. Thus, it is of great interest to study the stability conditions from anisotropic perturbations against the de Sitter expanding space during the very early cosmological epochs. A general analysis of the stability conditions for the effective theories of gravity such as the inflationary potentials in the flat FLRW Universe are very useful to characterize the  (in)stability of the physical models \cite{Dobado:1993a,Dobado:1995,Kao:1991,Kao:2001a,Kao:2001b,Kao:2010q,Kanti:1999, Maeda1989a,Walda1983,Barrow:2006aa,Barrow:2006bb}.

For the sake of completeness, we recall that the stability analysis of the non-redundant field equations in the Bianchi type I Universe has been studied in the isotropic limit \cite{Kao:2001b} and for anisotropic brane cosmology \cite{Kao:2002s}. For the Bianchi type I isotropic brane cosmology  \cite{Kao:2001b}, it was shown that any unstable mode of the isotropic perturbation with respect to a de Sitter background is also unstable with respect to such anisotropic perturbations. Furthermore, the anisotropic brane cosmology \cite{Kao:2002s} shows for both theories that the anisotropic expansion is dynamically smeared out in the long-time limit. This is not depending on the different types of equations of state. The Bianchi type I anisotropic cosmology is stable against any anisotropic perturbation for a perfect fluid or a dilaton field \cite{Kao:2002s}. In addition, the stability analysis \cite{Harko:2001s, Kao:2001cx, Dobado:1995, Maroto:1997se, Kao:200fw} indicates that all models in both theories are stable against any anisotropic perturbation. The condition of perturbatively (un)stable cosmological model is fulfilled when the modes ($\gamma$) of plane-wave equations are determined; $\gamma_+>0$ (unstable) or $\gamma_-<0$ (stable). Other examples can be taken from \cite{injp1}, for instance.

The instability of the static Einstein Universe in presence of quantum fluctuations has a particular relevance, especially at infinitely long time. Furthermore, it was found that the static Einstein Universe is unstable with respect to small radial perturbations \cite{Audrey, Mulryne:2005s, Guendelman:2011, Wu:2010fg}. It was argued that even when the cosmological models are perfectly fine-tuned to describe the early stages of the Universe, the quantum fluctuations - among others - generate inflationary expansion or even derive the Universe towards collapse at infinite time (age!) \cite{Mulryne:2005s,Guendelman:2011,Wu:2010fg}. 

The present paper is organized as follows. In Sec. \ref{sec:model}, we implement the FLRW metric to the inflationary era taking into account quantum potential corrections \cite{Das:2014}. For different equations of state (EoS), the perturbative (in)stability of the FLRW cosmology shall be elaborated in Sec. \ref{stablUNI}. Section \ref{conc} is devoted to the conclusions.

\section{Cosmological models with inflation field}
\label{sec:model}

We start from the FLRW metric \cite{Kao:1991}, 
\bea
ds^2 = - b^2(t)\, dt^2 + a^2(t)\, \left( \frac{dr^2}{1-\kappa r^2} + r^2\, d\Omega\right),  
\eea
where $\kappa$ is the curvature constant, $0$, $\pm1$ for a flat or closed or open Universe, respectively, $a(t)$ is the cosmic scale factor and $b(t)$ is the lapse function. For a perfect fluid with vanishing viscosity, the energy-momentum tensor reads 
\bea
T_{\mu \nu} &=& (\rho + p) \, u_{\mu}\, u_{\nu} + g_{\mu \nu}\, p,
\eea
where  $\rho$ is the comoving energy density, $p$ is the pressure, and $u_\mu$ is the four-velocity field. The subscripts $\mu$ and $\nu$ run over $0,1 \cdots, 3$. The energy-momentum conservation condition $D_\mu \, T^{\mu \nu}=0$ is apparently equivalent to the time evolution of energy density which defines the continuity equation,
\bea
\dot{\rho} &=& -3 \rho (1+\omega) H,
\eea
where $\omega$ characterizes effective barotropic equation-of-state, $\partial p=\omega\, c^2\,\partial\rho$ \cite{Tawfik:2012ty}. In the physical units, $\omega$ is given in $c_s^2/c^2$-units; speed of sound relative to speed of light through the medium of interest, e.g. cosmic background.

Second, when imposing a scalar field ($\phi$), the energy and pressure density, respectively, can be given as 
\bea
\rho_\phi &=& \frac{1}{2} \dot{\phi}^2 +V(\phi),\\
p_\phi &=& \frac{1}{2} \dot{\phi}^2 -V(\phi), \label{pressure0}
\eea
where the cosmological constant is assumed to be included in the inflationary potential $V(\phi)$. 

Third, the resulting equation of state  - in natural units - can be given as $\omega=(p+p_\phi)/(\rho+\rho_\phi)$, where the contributions from the inflationary scalar potential are taken into consideration. It is conjectured that no correlations (interactions) exist between the cosmic fluid and the inflationary potential field. 

It has been assumed that replacing the classical trajectories or geodesics by their quantum counterparts replaces the classical velocity field used in the Raychaudhuri equation by a quantum velocity field and thus constructs the so-callled quantum Raychaudhuri equation (QRE) \cite{Das:2014}. Consequently, QRE is believed to prevent the formation of singular points \cite{Ali:2015}. This doesn't show that the spacetime singularities are inevitable. Furthermore, one should remark that the second order Friedmann equation {\it naturally} contains quantum corrections, especially in the quantum mechanical description of the physical Universe. It was concluded that the quantum correction  \cite{Ali:2015} regardless its incorrectness \cite{Lashin:2015} makes the past singularity infinite and predicts an everlasting Universe.  The construction of QRE based on de Broglie-Bohm theory was sharply criticized \cite{Lashin:2016}.

The present work focuses on a systematic analysis for the perturbative (in)stability of both standard FLRW Universe and that the cosmology from quantum potentials \cite{Ali:2015}, especially in the inflationary era. In doing this, we follow the Raychaudhuri field equation back to the inflationary era and apply perturbative instability tests. 

Respectively, both Friedmann and Raychaudhuri field equations can be expressed as
\bea
H^2	&=& \frac{8\pi G}{3} \left[\rho+ \frac{1}{2}\dot{\phi}^2 +V(\phi) \right] - \frac{\kappa}{a^2}, \label{Freidemna} \\
\frac{\ddot{a}}{a} &=&-\frac{8\pi G}{3}\left[\frac{1}{2}(1+3\omega)\rho +\dot{\phi}^2 - V(\phi) \right], \label{Rary0} 
\eea
with $H=\dot{a}/a$ is the Hubble parameter. One can imply that
\bea
\dot{H} &=& - 4 \pi G \left[\dot{\phi}^2 + (1+\omega)\rho \right] + \frac{\kappa}{a^2}. \label{RayScalar}
\eea
This is the Raychaudhuri field equation in which the scalar field ($\phi$) and energy density ($\rho$) for different equations of state, are included.

For a non-symmetry reduced system of gravity, which is non-minimally coupled to the massive scalar
field, the action of the FLRW spacetimes gets modified  \cite{JCAP:Artymowski}
\bea
\mathcal{S} &=& \frac{1}{8\pi G} \int d^4x\, a^3 \left[ -6UH^2-6HU' \dot{\phi} + \frac{\dot{\phi}^2}{2} -V(\phi) \right],
\eea
where $U'(\phi) \equiv \partial_\phi U(\phi) $. The limit of minimally coupled scalar field is defined at $U(\phi)=1/2$.  The differentiation of this action with respect to $a$ and $\phi$, respectively, results in equations of motion 
\bea
\ddot{\phi} + 3 H \dot{\phi}+V'(\phi)&=& 6 U' \left[ \frac{\ddot{a}}{a}+H^2 \right], \label{eq:eom1} \\
2 U' \ddot{\phi} + 2 U \left[\frac{2\ddot{a}}{a} + H^2 \right]+ 2 U'' \dot{\phi}^2 + 4HU' \dot{\phi} &=& -8\pi G p. \label{eq:eom2}
\eea

Besides the scale factor, the determination of the scalar field canonical momenta helps in constructing the Hamiltonian ($\mathcal{H}$) \cite{JCAP:Artymowski}. Also, one can modify the second equation of motion, Eq. (\ref{eq:eom2}), through the cosmic term, such as
\bea
\ddot{\phi}+3 H \dot{\phi} &=& \frac{2 U' V - U V' -U' \dot{\phi}^2 (3 U''+\frac{1}{2})}{U+3 U'^2}. \label{EQMotion}
\eea
Two scalar fields which are non-minimally coupled to gravity, 
\bea
U(\phi) &=& \frac{1}{2}+ \frac{1}{2}\eta \phi^2, \label{pote0}\\
V(\phi) &=& \frac{\lambda}{4}\, \phi^4, \label{pote1}
\eea
can be introduced to the FLRW flat Universe. 

Equations (\ref{pote0}) and (\ref{pote1}) are conjectured to describe an inflationary era in early Universe, where $\eta,\, \lambda$ are constants. Furthermore, we assume that the scalar field very slowly varies so that the acceleration would be neglected, i.e. $\ddot{\phi}\ll 3H\dot{\phi}$ \cite{Liddle:2003w}. Then, Eq. (\ref{EQMotion}) can be written as
\bea
\eta\left(1+6 \eta\right)\;\phi \dot{\phi}^2 + 3H \left[1+\eta\left(1+6\eta\right) \phi^2 \right]\; \dot{\phi}+ \lambda \phi^3 = 0, \label{GenrEq0}
\eea
The time derivative of the inflation field ($\dot{\phi}$) leads to
\bea
\dot{\phi} &=& \frac{-3}{2\,\eta\phi} \left(1+\eta \phi^2\right) H \pm \left[\frac{3}{2\eta \phi} \left(1+\eta \phi^2\right)H - \frac{\lambda \phi^3}{3H (1+\eta \phi^2)}\right]. \label{GenralSol0}
\eea 
The general solution to the quadratic Eq. (\ref{GenrEq0}) is given in Eq. (\ref{GenralSol0}). For the sake of simplicity, let us take into consideration the positive sign. This leads to $\dot{\phi}= -\lambda \phi^3/\big[3H (1+\eta \phi^2)\big]$. 

In the given scalar field ($\phi$) and by substituting with the Friedmann equation, the Raychaudhuri field equation reads
\bea
\dot{H} &=&-\frac{3}{2} (1+\omega)\, H^2 -\frac{4 \pi G}{9} \left[\frac{\lambda \phi^3}{(1+\eta \phi^2)}\right]^2\, H^{-2}\nn \\ &-& \frac{6\, \epsilon_1\, \hbar^2}{m^2} (1+\omega) \left[6(1+\omega)^2 -\frac{81}{2}(1+w)+18\right]\, H^4, \label{CQRE}
\eea
where $\epsilon_1$ is constant and $m$ can be regarded as the smallest graviton (or axion) mass or the Planck mass. The subscript $1$ is added to merely distinguish it from the other slow-roll parameter; $\epsilon=-\dot{H}/H^2$.
It is obvious that the $\hbar^2$-term representing the quantum corrections vanishes in the limit ($\phi;\, \hbar \rightarrow 0$) \cite{Ali:2015}. 

It was claimed that Eq. (\ref{CQRE}) without the second term leads to an ageless Universe \cite{Ali:2015}. This proposal was critically commented \cite{Lashin:2015,Lashin:2016}, where the third term was corrected a\cite{Lashin:2015} and  both physical relevancy and mathematical correctness of the entire approach of QRE, which was based on Bohmian quantal trajectories were criticized.  For further details interested readers are kindly advised to consult Ref. \cite{Tawfik:2016biw}.

\begin{figure}[hbt]
\includegraphics[width=8.cm]{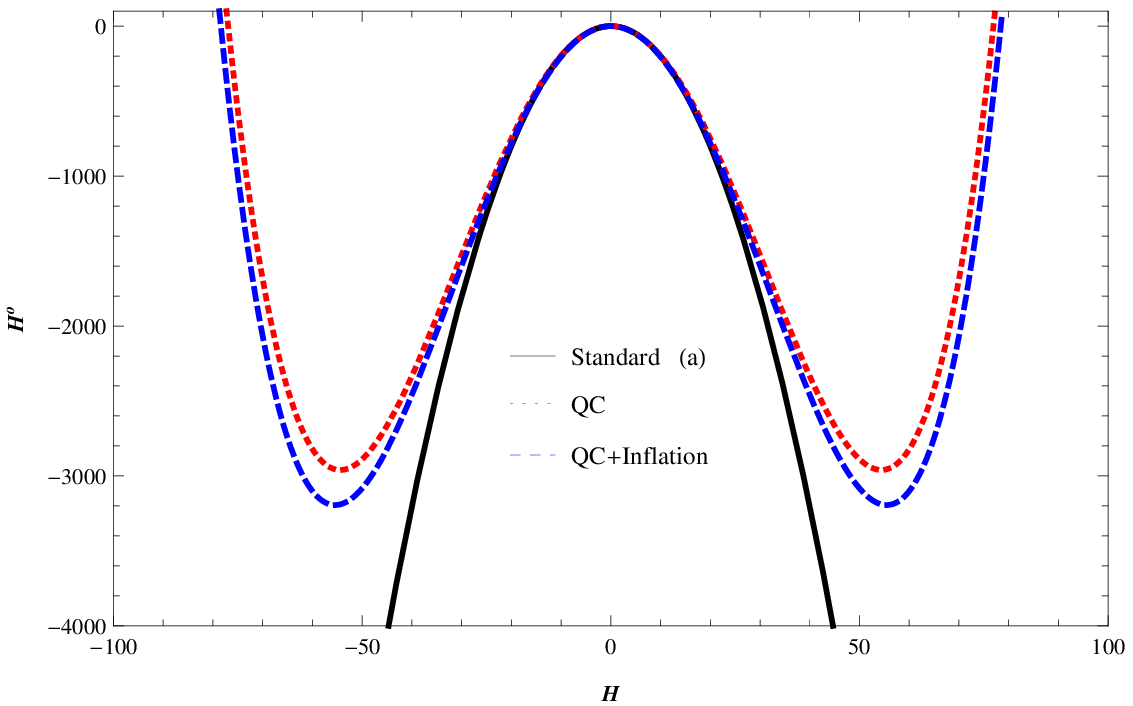}
\includegraphics[width=8.cm]{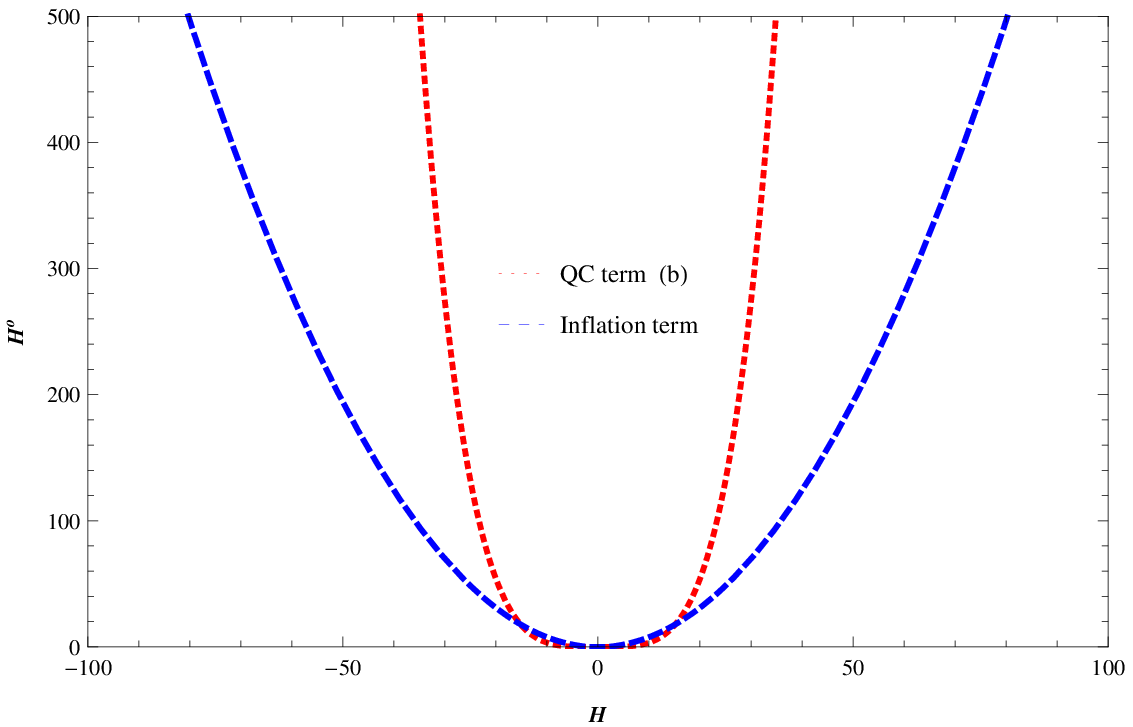}
\caption{Left-hand panel (a) depicts $\dot{H}$ as a function of $H$, Eq. (\ref{CQRE}), where the solid curve illustrates the first term in rhs, dotted curve stands for first and third terms (named as QC, quantum corrections), and dashed curve presents the three terms added together. Right-hand panel (b) shows the same but for second [named as Inflation] (dashed) and third term (dotted curve), separately. \label{fig:1}}
\end{figure}

In order to gain more insights from Eq. (\ref{CQRE}), we depict $\dot{H}$ vs. $H$ in Fig. \ref{fig:1}. In the left-hand panel (a), we show $\dot{H}$ as a function of $H$ as calculated from Eq. (\ref{CQRE}), where the solid curve represents the first term in rhs. This is the {\it standard} Raychaudhuri equation, in which the dependence of $\dot{H}$ on $H$ is well described by a quadratic relation describing parabola. Here, $\omega$ would play a role in determining whether the parabola can open up or down, especially for a hypothetical assumption that $\omega<-1$. The dotted curve depicts the results obtained from the first and the third terms. Here, we observe that at small $H$, the first term seems to become dominant, while the third one, the quantum corrections \cite{Ali:2015,Lashin:2015,Lashin:2016} turn to be dominant so that it inverts the direction where the parabola opens. 

The dashed curve illustrates the calculations based on adding up the three terms together. Here, we can estimate the similarities between the contributions added by the second and the third terms. While the third term depends on $\omega$, the second term doesn't do. The right-hand panel gives details about second and third term, separately.  

It should noticed that the calculations are very sensitive to the parameters in front of the third term. Unfortunately, the authors of \cite{Ali:2015,Lashin:2015} gave almost no hits about their parameters or how they performed their calculations. For our calculations we have assumed that $\omega=1/3$ and $\epsilon_1 (\hbar/m)^2 \simeq 1.667 \times 10^{-6}$ \cite{Tawfik:2016biw}. For the inflation term, the second term in rhs, we assume that the inflation field is given as $H^2/3m$, with $\lambda=1/2$ and $\eta=\sqrt{\lambda}$.

In hypothetical cases, such as Chaplygin gas, e.g. $\omega<-1$, one would expect that $\dot{H}$ always increases with increasing $H$, i.e. the resulting parabola opens up, especially that the second term always results is positive $\dot{H}$ with the increase in $H$, as well. The $H$-value, where $\dot{H}$ starts to increase with $H$ might taken as an upper limit determining the validity of the quantum corrections, the third term.

A few comments on Eq. (\ref{CQRE}) is now in order. First, the right-hand side combines quantum mechanics, e.g. $\hbar$, with  purely classical FLRW equation! Second, we didn't approve such an {\it ad hoc} imposed mixture. Third, the scope of this script is not to approve or disapprove the model suggested in Ref. \cite{Ali:2015}. This was carried out in literature  \cite{Lashin:2015,Lashin:2016,Tawfik:2016biw}. Fourth, the nonchalant introduction of quantum aspects based on Bohmian mechanics to the FLRW equation was superficially thought to determine quantum corrections. Fifth, Bohmian mechanics also known as de Broglie-Bohm theory gives an alternative (causal) interpretation of quantum mechanics, e.g. hidden variables interpretation accompanied by inevitability of nonlocality. On the other hand, Bohmian mechanics introduces a radically different perception of the underlying processes. Last but not least, the present script points out that Eq. (\ref{CQRE}) leads to an unstable Universe and therefore nonphysical and/or even improper.

\section{Results on perturbative (in)stability} 
\label{stablUNI}

The present article is devoted to determining of the (in)stability conditions of the proposed cosmology from quantum corrections. We first consider homogeneous scalar perturbations to the FLRW Universe with quantum corrections in presence of the scalar field $\phi$ as given in Eq. (\ref{CQRE}). Then, we determine their (in)stability modes.

\subsection{Perturbative instability of FRLW equation}

Perturbations to the Hubble parameter ($H$), confined energy density ($\rho$), pressure ($p$), equation of state ($\omega$), and the scalar field ($\phi$), respectively, can be expressed as
\bea 
\dot{H} &=& \bar{\dot{H}}\,+\, \delta\, \dot{H}(t,x),\label{pert01} \\
\rho &=& \bar{\rho} + \delta\, \rho(t,x),\label{pert02} \\
p &=& \bar{p} + \delta\, p(t,x), \label{pert03} \\
\omega &=& \bar{\omega} + \delta\, \omega(t,x), \label{pert04}\\ 
\phi &=& \bar{\phi} + \delta\, \phi(t,x),
\label{pert0}
\eea
where bars donate a spatial average.  It is noteworthy highlighting that the special and temporal dimensions in an expanding Universe are mutually depending on each other. For simplicity, we focus on time-independent special perturbation. 
\begin{itemize}
\item Firstly, let us assume that the $H^2$-coefficient in the first term of Eq. (\ref{CQRE}) is given as $\alpha=-\frac{3}{2} (1+\omega)$. Then, the perturbation to this term leads to  
\bea
\alpha + \delta \alpha &=& -\frac{3}{2} (1+\bar{\omega} + \delta\, \omega).
\eea 
When eliminating the higher orders, we get that  
\begin{eqnarray}
\delta \alpha &=&  -\frac{3}{2} \delta \omega = -\frac{3}{2} \left[\left(\omega+\delta\omega\right)- \omega\right] =  -\frac{3}{2} \left[ \frac{p+\delta p}{\rho+\delta} -\frac{p}{\rho} \right] =-\frac{3}{2}\frac{p}{\rho} \left[ \frac{1+ \left(\frac{\delta p}{p}\right)}{1+ \left(\frac{\delta \rho}{\rho}\right)} -1 \right] \nonumber \\  &=&  -\frac{3}{2}\frac{p}{\rho} \left[ \left( 1+ \frac{\delta p}{p}\right) \left(1+ \frac{\delta \rho}{\rho}\right)^{-1} -1 \right] =  -\frac{3}{2}\frac{p}{\rho} \left\{\left[1+\frac{\delta p}{p}-\frac{\delta \rho}{\rho} + {\cal O}(\delta^2)\right]-1\right\}. \label{Exp1}
\end{eqnarray}
For $\delta p\ll \delta\rho \ll 1$, $\delta \alpha = -\frac{3}{2} \frac{p}{\rho} \left(-\delta \rho/\rho\right) = \frac{3}{2} \omega \delta$, where $\delta = \delta \rho/\rho$.

\item Secondly, we assume that the correction due to the scalar field ($\phi$) is given as
\bea
\sigma_\phi  &=&  - \frac{4\pi G}{9} \left[\frac{\lambda \phi^3}{(1+\eta \phi^2)}\right]^2 = -\frac{4\pi G}{9} \lambda^2  \phi^6(1 - 2 \eta \phi^2).
\eea
The perturbation to this quantity can be expressed as
\bea
 \delta \sigma_\phi =-\frac{8\pi G}{9} \lambda^2 \phi^6 (3 - 8 \eta \phi^2) \delta',
\eea
where $\delta' \equiv \delta\phi/\phi$.
\item Finally, we assume that the third term of Eq. (\ref{CQRE}), quantum potential correction, can be named as
\bea
\xi &=& - \frac{6 \epsilon_2 \hbar^2}{m^2}\, (1+w) \left[6(1+w)^2 -\, \frac{81}{2}(1+w)+18\right], \label{eq:qcorrALI}
\eea
where $\epsilon_2$ is an arbitrary constant. 
Then, by applying perturbation as done in Eq. (\ref{pert0}), the first-order perturbation to $\omega$ reads
\bea
\delta \xi = - 270 \frac{\epsilon_2 \hbar^2}{m^2} \, \omega \delta. \label{dxi}
\eea
The first-order perturbations to the FLRW Raychaudhuri equation can be given as
\bea
\dot{\delta H} & = & \left(2 \alpha H + 4 \xi H^3 -2\sigma_\phi  H^{-3}\right)\, \delta H+\left(\delta \alpha\right)\, H^2 +\left(\delta \sigma_\phi\right)\, H^{-2} + \left(\delta \xi\right)\, H^4. \label{Rey3inflationa}
 \eea
By substituting the parameters $\alpha$, $\xi$, and $\sigma_\phi$ and their perturbations into Eq. (\ref{Rey3inflationa}), we obtain
\bea 
\dot{\delta H} &=& \left\{-3(1+w)H-\frac{24 \epsilon_1\hbar^2}{m^2}(1+w) \left[6(1+w)^2 - \frac{81}{2}(1+w)+ 18\right] H^3  \nn \right. \\ &&\left. +\frac{8\pi G}{9} \lambda^2 \phi^6(1 -2\eta \phi^2) H^{-3} \right\} \delta H \,+\, \left\{\frac{3\omega}{2} H^2  - 270  \frac{\epsilon_1 \hbar^2 \omega}{m^2} H^4  \right\} \delta \nn \\ &-& \left\{ \frac{8\pi G}{9} \lambda^2 \phi^6 (3 - 8 \eta \phi^2) H^{-2} \right\} \delta ' \label{HInfala0}. 
\eea
The time derivatives of $\alpha$ and $\dot{\alpha'}$ can respectively be given as
\bea
\dot{\delta} & = & -3(1+\omega) \delta H \,-3 H\omega\,  \delta, \label{deldot0} \\ 
\dot{\delta '} &=& \frac{\lambda \phi^2}{3}(1-\eta\phi^2)H^{-2} \delta H - \frac{2\lambda \phi^2}{3}(1-2\eta\phi^2)H^{-1} \delta '.  \label{alpha02}
\eea
\end{itemize}

It obvious that in the limit $\phi;\, \hbar \rightarrow 0$, the first-order perturbations in the FLRW Raychaudhuri equation, Eq. (\ref{CQRE}), are restored  
\bea
 \delta \dot{H} &=& -3 \left(1+\omega\right) H \delta H + \frac{3}{2} H^2 \omega \delta. \label{delH} 
\eea
Let us assume that $\dot{\delta} = \partial/\partial t \left(\delta \rho/\rho\right)$. Accordingly, from the coupling between Eqs. (\ref{deldot0}) and (\ref{delH}), it becomes straightforward to determine the second-time derivative of the homogeneous Eq. (\ref{delH}),
\bea
A\, \ddot{\delta H} + B\, \dot{\delta H} + C \, \delta H  = 0, \label{sol1}
\eea
where $A=1$, $B=-3 (2+3\omega) H$ and $C= (-9/2)(1+ 5\omega) H^2$. The general solution to this homogeneous ordinary-differential equation reads
\bea
\delta H (t) &=& \beta_1 \exp{\left[ \gamma_+ \,t\right]}\,+\, \beta_2 \exp{\left[ \gamma_- \,t\right]}, \label{simpSolA}
\eea
where $\beta_1$ and $\beta_2$ are integration constants and  the modes $\gamma_{\pm}$ are assumed as zeroth order perturbation, i.e., $H\equiv H_0$  
\bea
\gamma_{\pm} &=& -B\pm \sqrt{B^2\,-4\,A\,C}/2A =\frac{3}{2} \left[ (2+3\omega) \pm \sqrt{6+22\omega + 9\omega^2  }\right] H_0. \label{simpSolA}
\eea
The presence of a finite cosmological constant ($\Lambda \neq 0$) converts Eq. (\ref{sol1}) to an inhomogeneous ordinary differential equation. A few comments on the stability modes ($\gamma_{\pm}$) is now in order.
\begin{itemize}
\item The square root is less than the first two terms, $3 (2+3\omega)$. This means that this solution is apparently identical to a stable equation in presence of an inflationary era of the de Sitter solution \cite{Dobado:1995,Kao:2001a,Kao:2001b}. 
\item Occasionally, the square root might possess instability modes, i.e. $\gamma_+ >,0$. In this case, despite the assumption that the inflationary era will come to an end, rapidly, so that such an unstable mode takes place. This likely sharpens the stability of the isotropic space \cite{Kao:2001a,Kao:2001b}. It has been concluded that even if such an unstable mode for the de Sitter perturbation would happen, it will be unstable against the anisotropic perturbation.
\end{itemize}

\subsection{Impacts of various equations of state}

In this section, we explore the specific states of the inflationary era and intend to obtain some additional restricting conditions on having stable FLRW Universe, especially during the inflation, in which different equations of state (EoS) are conjectured to characterize the cosmic background geometry. 
\begin{itemize}
\item{Radiation dominated era:} EoS is characterized by $\omega=1/3$ and thus
\bea
-\frac{1}{2} \left(-9+\sqrt{129}\right)\, H_0 < \gamma < \frac{1}{2} \left(-9+\sqrt{129}\right)\, H_0,
\eea
which refers to unstable and stable modes during the inflationary  era of the flat FLRW Universe, see previous section.  This inequality restricts the lowest order perturbation of the Hubble parameter ($H_0$) in the flat FLRW Universe. We notice that the right-hand side is always negative as $H_0$ is a positive quantity, i.e. stable mode and isotropic perturbation are obtained, $\gamma_-<0$. It is obvious that the left-hand side is positive $\gamma_+>0$, i.e., unstable modes and anisotropic perturbation.
\item{Matter energy dominated era:} $\omega=0$, 
\bea
-\frac{3}{2} \left(-2+\sqrt{6}\right)\, H_0 < \gamma < \frac{3}{2} \left(-2+\sqrt{6}\right)\, H_0,
\eea
again this inequality reveals both unstable and stable modes. As discussed in the radiation dominated era, both modes are likely possible 
\item{Vacuum energy dominated era:} During inflation and accelerating expansion, the EoS is characterized by negative $\omega$, for instance,  $-1<\omega<-1/3$, 
\bea
\omega &=&-1,  \qquad   \frac{3}{2} \left(-1- i \sqrt{7}\right)\, H_0 < \gamma < \frac{3}{2} \left(-1+i \sqrt{7}\right)\, H_0, \label{inequal_1} \\
\omega &=-&1/3, \qquad  \frac{3}{2} \left(1-\frac{i}{\sqrt{3}}\right)\, H_0 < \gamma < \frac{3}{2} \left(1+\frac{i}{\sqrt{3}}\right)\, H_0.\label{inequal_2}
\eea
At $\omega<−1$, the dark energy density increases due to the Universe expansion \cite{Hogan:2007a}. For the sake of completeness, we summarize that the dark energy density slowly decreases as the Universe expands. It is noteworthy recalling that the standard FLRW Universe at vanishing $\Lambda$ and  $-1<\omega<-1/3$ is unstable against a small perturbation in both solutions (inequalities), Eqs. (\ref{inequal_1}) and (\ref{inequal_2}). 
\end{itemize}

So far, the scalar perturbation leads to (un)stable modes of the flat FLRW Universe, especially during the inflationary era. The stable modes are characterized by positive EoS, i.e. $\omega = 1/3$ and $\omega=0$ for radiation and matter dominated era, respectively. On the other hand, the vacuum-energy-dominating era is characterized by negative $\omega$, that might be, for instance, ranging between $-1$ and $-1/3$. In this case, we have obtained that the flat FLRW Universe without cosmological constant and coupled to a scalar field becomes unstable against scalar perturbation.

\subsection{Perturbative instability of Raychaudhuri equation from quantum potential}

Equations (\ref{HInfala0}), (\ref{deldot0}) and (\ref{alpha02}), can be respectively rewritten as
\bea
(\mathbf{D} - x_1) \delta H - x_2 \delta - x_3 \delta ' &=& 0, \label{HInfala0I}\\
-y_1  \delta H + (\mathbf{D} - y_2)  \delta +y_3 \delta ' &=& 0, \label{deldot0I} \\
-z_1  \delta H + z_2 \delta  + (\mathbf{D} - z_3) \delta ' &=& 0,\label{alpha02I}
\eea
where $\mathbf{D}\equiv d/dt$ and the coefficients 
\bea
x_1 &=& -3(1+w)H-\frac{24 \epsilon_1\hbar^2}{m^2}(1+w) \left[6(1+w)^2 - \frac{81}{2}(1+w)+ 18\right] H^3 \nn \\
&+& \frac{8\pi G}{9} \lambda^2 \phi^6(1 -2\eta \phi^2) H^{-3}, \label{eq:xx1}\\
x_2 &=& \Big( \frac{3}{2} H^2  - 270 \frac{\epsilon_1 \hbar^2}{m^2} H^4\Big)\, \omega, \label{eq:xx1}\\
x_3 &=& -\frac{8\pi G}{9} \lambda^2 \phi^6 \left(3 - 8 \eta \phi^2\right) H^{-2}, \label{eq:xx3}\nn \\
y_1 &=& \lambda_3 = -3 (1+\omega), \label{eq:yy1} \\
y_2  &=& -\lambda_4 = 3 H\omega, \label{eq:yy2} \\
y_3 &=& 0, \label{eq:yy3}\nn \\
z_1 &=& \frac{\lambda \phi^2}{3}\left(1-\eta\phi^2\right)\, H^{-2}, \label{eq:zz1}\\
z_2 &=& 0, \label{eq:zz2}\\
z_3 &=&  - \frac{2\lambda \phi^2}{3}\left(1-2\eta\phi^2\right)\, H^{-1}. \label{eq:zz3}
\eea

The set of linear differential equations (\ref{HInfala0I}), (\ref{deldot0I}), and (\ref{alpha02I}) can be combined  to a third-order differential equation in $\delta H$, 
\bea
\left[\mathbf{D}^3 - (x_1+y_2+z_3) \mathbf{D}^2 +(x_1\, y_2 + x_1\, z_3 + y_2\, z_3 - x_2\, y_1 - x_3\, z_1) \mathbf{D} \right. &-&  \nn \\ \left. (x_1\, y_2\, z_3 - x_2\, y_1\, z_3 -x_3\, y_2\, z_1)\right]  \delta H &=& 0, \label{th3order}
\eea
where $\mathbf{D}^n$ stands for $n$-th time derivative.  Equation (\ref{th3order}) can be rewritten as
\bea
\left[ (\mathbf{D}- x_1 -\Omega)\; (\mathbf{D}- y_2 +\Omega) \;(\mathbf{D}- z_3) \right]\, \delta H = 0. \label{th3orderForm}
\eea
It is obvious that $\Omega$ can be determined by comparing Eqs. (\ref{th3order}) and  (\ref{th3orderForm}) with each other,
\bea
 (x_1-y_2+z_3)  \Omega + \Omega^2 &=& x_2 y_1 + x_3 z_1, \label{II0}\\ 
 (x_1-y_2) z_3 \Omega + z_3 \Omega^2 &=&x_2 y_1 z_3 +  x_3 z_1 y_2, \label{II1}
\eea 
By multiplying both sides of Eq. (\ref{II0}) by $z_3$ and then subtracting it from Eq. (\ref{II1}),  we obtain that
\bea
\Omega &=&  x_3 z_1 \frac{z_3 - y_2}{z_3^2}.
\eea 
From Eqs. (\ref{eq:xx3}), (\ref{eq:yy2}), (\ref{eq:zz1}), and (\ref{eq:zz3}), we get 
\bea
\Omega &\simeq & 2\, \pi\, \lambda\, G\, \omega\, \phi^4 \left[3+\frac{2\, \lambda}{3\, \omega}\, \phi^2 \left(1-2\, \eta\, \phi^2\right)\, H^{-2}\right]\, H^{-1}.
\eea
Hence, the general solution to Eq. (\ref{th3orderForm}) can be defined as
\bea
\delta H &=& \mathcal{A}_1\, e^{\psi_1\; t} +\mathcal{A}_2\, e^{\psi_2\; t} +\mathcal{A}_3 \,e^{\psi_3\; t} \label{solutionQC}
\eea 
where $\mathcal{A}_1$, $\mathcal{A}_2$ and $\mathcal{A}_3$ are normalization constants and
\bea
\psi_1 = x_1+\Omega &=& -3(1+w)H-\frac{24 \epsilon_1\hbar^2}{m^2}(1+w) \left[6(1+w)^2 - \frac{81}{2}(1+w)+ 18\right] H^3 \nn \\ &+& \frac{8\pi G}{9} \lambda^2 \phi^6(1 -2\eta \phi^2) H^{-3} + 2 \pi G \lambda \omega  \phi^4 \left(3+ \frac{2\lambda \phi^2}{3\omega}(1-2\eta\phi^2) H^{-2} \right) H^{-1}, \label{1stterm} \\
\psi_2 = y_2-\Omega &=& 3 H \omega -  2 \pi G \lambda \omega  \phi^4 \left(3+ \frac{2 \lambda \phi^2}{3\omega}(1-2\eta\phi^2) H^{-2} \right) H^{-1}, \label{2ndterm}\\
\psi_3 = z_3 &=&  \frac{-2\lambda \phi^2}{3} (1-2\eta\phi^2) H^{-1}.  \label{3rdterm}
\eea

\subsection{Choice of parameters}

As introduced in earlier sections, the numerical estimation for the (un)stable modes plays a crucial role in our determination of the cosmological (in)stability. The analysis of the perturbative (in)stability of the FLRW Universe from quantum potential, especially during the inflationary era is strongly depending on the choices of the various parameters. The parameter obtained with the solution, Eq. (\ref{solutionQC}), and the inflation potentials, Eqs. (\ref{pote0}) and (\ref{pote1}), can be determined when assuming that $\epsilon_1$, $\hbar$ and the mass ($m$) have the values $1/6$, $4.135\times10^{-15}~$eV s, and $\sim 10^{-32}~$eV/c$^2$, respectively \cite{Ali:2015}.  Furthermore, the relation between the coupling constants $\eta$ and $\lambda$ is given by the normalization of primordial inhomogeneities. Several different values have been considered;  $\eta_n = 4.7 \times 10^n \sqrt{\lambda};\, n=0,\,1,\cdots$. In  our calculations, we assumed that $\eta =47000 \sqrt{\lambda}$, in order to run analysis for various coupling constant $\lambda$ limits within the allowed range of the mass ($m$). 

For instance, in case of the Higgs scalar field \cite{ezrukov2009} with an uncertainty of the order of $2~$GeV, it was found that the 2012-results by CMS \cite{CMS2012} and from ATLAS \cite{ATLAS2012} collaborations are consistent with the standard model for Higgs inflation. However, most recent results from both experiments \cite{CMS2012, ATLAS2012} suggest that $\lambda$ should not be too big. Since the mass of the Higgs is close to the minimally allowed value, which does not violate electroweak vacuum, we set $\lambda$ to $1/2$, which is the same order of magnitude as in the standard model. 

It is noteworthy highlighting that the authors of Ref. \cite{Ali:2015} did not elaborate which parameters choices they made! This was discussed in Refs. \cite{Lashin:2015,Tawfik:2016biw}, where the remarkable impacts of the parameters choices have been reported.

\subsection{Impacts of various equations of state}

As done with the FLRW equation, we shall summarize in the following how the (un)stable modes of Raychaudhuri equation from quantum potential, Eq. (\ref{CQRE}), vary with different equations of state including matter-/radiation-dominated eras (positive $\omega$) and the dark energy and the cosmological constant (negative $\omega$).

\begin{itemize}

\item Radiation dominated era, i.e. $\omega=1/3$, leads to unstable and stable modes for Eq. (\ref{solutionQC}), in which the flat FLRW cosmology, the inflation scalar field ($\phi$), and the quantum potential corrections are assumed. However, in Eq. (\ref{1stterm}), the bracket term ($\psi_1$) is always negative, while the other terms ($\psi_2$ and $\psi_3$), Eqs. (\ref{2ndterm}) and (\ref{3rdterm}), respectively, are positive quantities at all positive values of the time ($t$), the Hubble parameter ($H$), and the inflation potential field ($\phi$). Therefore, the solution with negative $\psi_1<0$ apparently refers to a stable mode and isotropic de Sitter space, while positive $\psi_2$ and $\psi_3$ refer to unstable modes and anisotropic perturbation against scalar perturbation of Eqs. (\ref{pert01}), (\ref{pert02}), (\ref{pert03}), (\ref{pert04}), and (\ref{pert0}).

\item Matter energy dominated era, i.e. $\omega=0$,  results in an unstable mode and an anisotropy perturbation. This EoS plays an essential role in the definition of $\psi_1$ and  $\psi_2$, Eqs. (\ref{1stterm}) and (\ref{2ndterm}), respectively. Both terms tend to infinity, while $\psi_3$ in Eq. (\ref{3rdterm}) is always positive for all positive values of the time ($t$), the Hubble parameter ($H$) and the inflation potential field ($\phi$).

\item Vacuum energy dominated era, for instance, dark energy or finite cosmological constant, for instance $-1<\omega<-1/3$, leads to the same results as that obtained in the radiation dominated era. Stable modes and isotropic de Sitter space which are resulted from perturbation apparently appears in $\psi_1$, Eq. (\ref{1stterm}), i.e. negative $\psi_1<0$. Positive $\psi_2$ and  $\psi_3$, Eqs. (\ref{2ndterm}) and (\ref{3rdterm}), respectively, lead to unstable modes and anisotropic de Sitter cosmology from quantum potential corrections.
\end{itemize}

\section{Conclusions} 
\label{conc}

The choices of many dependent variables characterizing the early Universe such as the  curvature constant, form of the manifold, the equation of state, the coupling constant, etc. are very crucial for securing a stable de Sitter background. It has been shown that the evolution of a small perturbation against isotropic and anisotropic FLRW background space helps in distinguishing between stable and unstable modes. The stable modes guarantee stability and also ensure anisotropy of the de Sitter space. On the other hand, the unstable mode indicates that the isotropic background is unstable against any small anisotropic perturbation. Therefore, only a small anisotropy in the early Universe could be generated by an arbitrary small anisotropic perturbation.

By replacing the classical trajectories by their quantum (Bohmian) counterparts, modifications to the second-order Friedmann equation have been reported \cite{Ali:2015}. This was nothing but the Raychaudhuri equation with a quantum correction term.  It was wrongly argued that this correction term avoids the big-bang singularity.  Furthermore, it was proposed that this correction term leads to an everlasting Universe. Critical comments on both conclusions and even on the correctness of the key expressions of this work have been discussed in literature \cite{Lashin:2015}. 
The physical relevancy and even the mathematical correctness of QRE, which was based on Bohmian quantal trajectories, were critically commented \cite{Lashin:2016}.

Instead of proposing further corrections, we have - in an early work - analyzed the perturbative (in)stability conditions \cite{Tawfik:2016biw}. We concluded that the quantum potential corrections and their additional parameters ($\epsilon$ or $\epsilon_1$, $\hbar$, and $m$) obviously strengthen the perturbative instability of our Universe. The scope of the present and the previous work \cite{Tawfik:2016biw} is not {\it solving} the Big Bang singularity problem. The perturbative quantum corrections, as utilized in Ref. \cite{Ali:2015} and fairly criticized in Ref. \cite{Lashin:2015,Lashin:2016} are not able to solve the Big Bang singularity due to absence of dynamics. To this end, one should recall what was already done on the basis of Borde-Guth-Vilenkin theorem, which holds even beyond General Relativity for any background expanding in average.

In the present work, we have checked the perturbative instability during the inflationary era.  During this stage of the early Universe evolution, best conditions for possible quantum effects are very likely. We have found that both unstable and stable modes are incompatible with the ones which were obtained from the FLRW Universe, to which a scalar field is added, without quantum potential corrections. Furthermore, we have shown that in the inflationary era, an unstable mode for a small (an)isotropic perturbation does exist for different equations of state. We have derived inequalities for both unstable and stable solutions to the standard FLRW space. These inequalities reveal that the flat FLRW Universe with a flat curvature likely possesses both instability and stability modes. On the other hand, negative stability modes ($\gamma_-$), and isotropic FLRW background space have been obtained for the radiation- and matter-dominated eras, merely. But the stability modes exist in case of finite cosmological constant and also if the vacuum energy dominates the cosmic background.  

The present work proves that during the inflationary era both unstable and stable modes are not compatible with the results obtained from FLRW Universe. Thus, we conclude that the Raychaudhuri cosmology from quantum potential is unstable against a small perturbation.

\end{document}